\documentclass[structabstract]{aa}  
\usepackage{graphicx}
\usepackage{txfonts}
\usepackage{natbib}
\begin{document}
\title{Multiple low-turbulence starless cores associated with
  intermediate- to high-mass star formation\thanks{The fits-files of
    the continuum and line images are available in electronic form at
    the CDS via anonymous ftp to cdsarc.u-strasbg.fr (130.79.128.5) or
    via http://cdsweb.u-strasbg.fr/cgi-bin/qcat?J/A+A/.}}


   \author{H.~Beuther
          \and
          Th.~Henning
          }
   \institute{Max-Planck-Institute for Astronomy, K\"onigstuhl 17,
              69117 Heidelberg, Germany\\
              \email{beuther@mpia.de, henning@mpia.de}
             }


\abstract
{}
{Characterizing the gas and dust properties prior to and in the
  neighborhood of active intermediate- to high-mass star formation.}
{Two Infrared Dark Clouds (IRDCs) -- IRDC\,19175-4 and IRDC\,19175-5
  -- that are located in the vicinity of the luminous massive
  star-forming region IRAS\,19175+1357, but that remain absorption
  features up to 70\,$\mu$m wavelength, were observed with the Plateau
  de Bure Interferometer in the 3.23\,mm dust continuum as well as
  the N$_2$H$^+$(1--0) and $^{13}$CS(2--1) line emission.}
{While
  IRDC\,19175-4 is clearly detected in the 3.23\,mm continuum, the
  second source in the field, IRDC\,19175-5, is only barely observable
  above the $3\sigma$ continuum detection threshold.  However, the
  N$_2$H$^+$(1--0) observations reveal 17 separate sub-sources in the
  vicinity of the two IRDCs.  Most of them exhibit low levels of
  turbulence ($\Delta v \leq 1$\,km\,s$^{-1}$), indicating that the
  fragmentation process in these cores may be dominated by the
  interplay of thermal pressure and gravity, but not so much by
  turbulence.  Combining the small line widths with the non-detection
  up to 70\,$\mu$m and the absence of other signs of star formation
  activity, most of these 17 cores with masses between sub-solar to
  $\sim$10\,M$_{\odot}$ are likely still in a starless phase.  The
  N$_2$H$^+$ column density analysis indicates significant abundance
  variations between the cores. Furthermore, we find a large CS
  depletion factor of the order 100. Although the strongest line and
  continuum peak is close to virial equilibrium, its slightly broader
  line width compared to the other cores is consistent with it being in a
  contraction phase potentially at the verge of star formation.  Based
  on the 3.23\,mm upper limits, the other cores may be gravitationally
  stable or even transient structures. The relative peak velocities
  between neighboring cores are usually below 1\,km\,s$^{-1}$, and we
  do not identify streaming motions along the filamentary structures.
  Average densities are between $10^5$ and $10^6$\,cm$^{-3}$ (one to
  two orders of magnitude larger than for example in the Pipe nebula)
  implying relatively small Jeans-lengths that are consistent with the
  observed core separations of the order 5000\,AU.  Environmental
  reasons potentially determining these values are discussed.}
{These observations show that multiple low- to intermediate-mass
  low-turbulence starless cores can exist in the proximity of more
  turbulent active intermediate- to high-mass star-forming regions.
  While masses and levels of turbulence are consistent with low-mass
  starless core regions, other parameters like the densities or
  Jeans-lengths differ considerably. This may be due to environmental
  effects. The quest for high-mass starless cores prior to any star
  formation activity remains open.}

   \keywords{Stars: formation  --
                Stars: individual: IRDC\,19175-4, IRDC\,19175-5 --
                Techniques: interferometric --
                Line: formation --
                Line: profiles --
                Turbulence 
               }

   \maketitle

\section{Introduction}
\label{intro}

Although our understanding of massive star formation has improved
tremendously over the last decade, most progress was made toward
relatively evolved regions like High-Mass Protostellar Objects
(HMPOs), Hot Molecular Cores (HMCs) or Ultracompact H{\sc ii} regions
(UCH{\sc ii}s), see for example several recent reviews like
\citet{beuther2006b,bonnell2006,cesaroni2006,hoare2006,zinnecker2007}.
Toward even earlier evolutionary stages prior or at the onset of
massive star formation, in recent years significant progresses have
been made in identifying large samples of Infrared Dark Clouds (IRDCs)
via absorption studies in the mid-infrared with the MSX, ISO and
Spitzer satellites (e.g.,
\citealt{egan1998,bacmann2000,simon2006,jackson2008}). Such IRDCs are
expected to harbor genuine high-mass starless cores (HMSCs) as well as
high-mass cores with embedded low- to intermediate-mass protostars in
the early accretion phase (e.g.,
\citealt{forbrich2004,klein2005,sridharan2005,rathborne2005,rathborne2006,beuther2007a,beuther2007g}).
While the samples contain more than 10000 IRDCs, it is interesting to
note that from the published higher spatial resolution case studies
(of the order 20) none is a real HMSC but all have identified embedded
accreting protostars (e.g.,
\citealt{beuther2005d,beuther2007a,birkmann2006,ormel2005,pillai2006b,rathborne2005,rathborne2008,forbrich2009}).
Hence, a genuine HMSC has not yet been unambiguously identified in the
literature.

To overcome this obvious missing link, here we report
high-spatial-resolution observations and analysis of two excellent
HMSC candidates in the dust continuum and the N$_2$H$^+$ line emission
to investigate their physical properties prior to any star formation
activity. N$_2$H$^+$ is known to be strong in low-mass starless cores
(e.g., \citealt{bergin2002,tafalla2004,chen2007}) as well as massive
IRDCs with already embedded, very young protostars (e.g.,
\citealt{beuther2005d}). Furthermore, it is a good chemical tracer of
the cold early evolutionary phase because it does not deplete much
onto grains at low temperatures but it starts getting destroyed when
the core heats up releasing the CO from the dust grains (which was
frozen onto them during earlier cold phases) that subsequently reacts
with N$_2$H$^+$ in the gas-phase chemistry \citep{bergin1997}.  While
the dust continuum emission pinpoints the locations of the densest gas
and dust components, the line emission will be used to determine the
gas kinematics of the region.

The two HMSC candidate sources are located in the vicinity of the
High-Mass Protostellar Object (HMPO) IRAS\,19175+1357.  During our
initial study of the dust continuum emission of a sample of HMPOs
\citep{sridha,beuther2002a}, we serendipitously detected mm continuum
peaks offset from the IRAS sources that turned out by comparison with
the MSX database to be associated with IRDCs \citep{sridharan2005}.
The average NH$_3$ temperatures and line widths of these objects are
significantly smaller than those from typical HMPOs and UCH{\sc ii}s,
hence putting them at an earlier evolutionary stage. However,
follow-up SiO(2--1) observations revealed SiO line wing emission
toward $\sim$40\% of the sample, indicating that star formation has
started in at least that fraction of sources \citep{beuther2007g}.

\begin{figure}[htb] 
\centering
\includegraphics[angle=-90,width=0.48\textwidth]{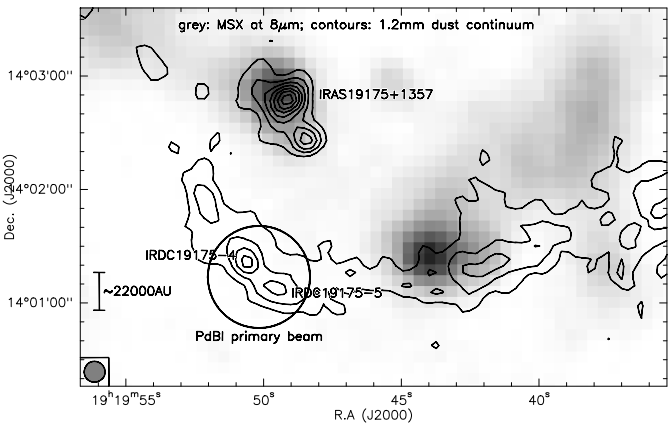}
\caption{Large-scale emission in the environment of the sources
  IRDC\,19175-4, IRDC\,19175-5 and IRAS\,19175+1357. The grey-scale
  shows the MSX 8\,$\mu$m mid-infrared emission, and the contours
  outline the 1.2\,mm continuum emission from \citet{beuther2002a} in
  steps of 20\,mJy\,beam$^{-1}$ ($3\sigma$
  rms$\sim$15\,mJy\,beam$^{-1}$). The circle presents the PdBI primary
  beam size of $\sim 54''$ encompassing the two target regions. A
  scale-bar and the IRAM 30\,m beam size of the 1.2\,mm continuum
  observations are shown at the bottom left.}
\label{overview}
\end{figure}

To identify a candidate HMSC for this sample with the given datasets,
we require the source to have (a) mm continuum emission, (b) no
detectable SiO emission, (c) narrow NH$_3$ and H$^{13}$CO$^+$ line
width, (d) low NH$_3$ temperatures and (e) no detections in the
Spitzer IRAC and MIPS 24 and 70\,$\mu$m bands. Here, we select from
the original sample two particularly promising HMSC candidates that
fulfill these criteria and even fit within one PdBI 3\,mm primary beam
(Fig.~\ref{overview}).

The two sources IRDC\,19175-4 and IRDC\,19175-5 with a projected
separation of $\sim$22000\,AU are part of the larger-scale filamentary
star-forming region associated with IRAS\,19175+1357
(Fig.~\ref{overview}) at an estimated distance of $\sim$1.1\,kpc
\citep{sridharan2005}. The luminosity of the main IRAS source in
Fig.~\ref{overview} is approximately $10^3$\,L$_{\odot}$, lower than
originally estimated based on a wrong distance in \citet{sridha}.
\citet{sridha} found weak cm continuum emission indicative of ionized
gas in the vicinity of the IRAS\,19175+1357 source. However, no maser
emission or strong CO line wings were detected. The latter indicates
either only weak or no outflow activity anymore toward this main
source, or the outflow could be oriented close to the plane of the
sky. Since the $v_{\rm{lsr}}$ of the main IRAS source and the two
IRDCs are approximately the same (6.9 versus 7.7\,km\,s$^{-1}$), we
can safely assume that both regions are at about the same distance.
Hence their projected separation of $\sim$0.45\,pc is a good
approximation of their real separation.  Although we cannot proof that
the IRAS source directly influences the two IRDCs, both regions should
nevertheless stem from the same larger gas clump that originally
fragmented into the already more evolved IRAS source and the younger
IRDCs.

\begin{figure*}[ht] 
\centering
\includegraphics[angle=-90,width=0.98\textwidth]{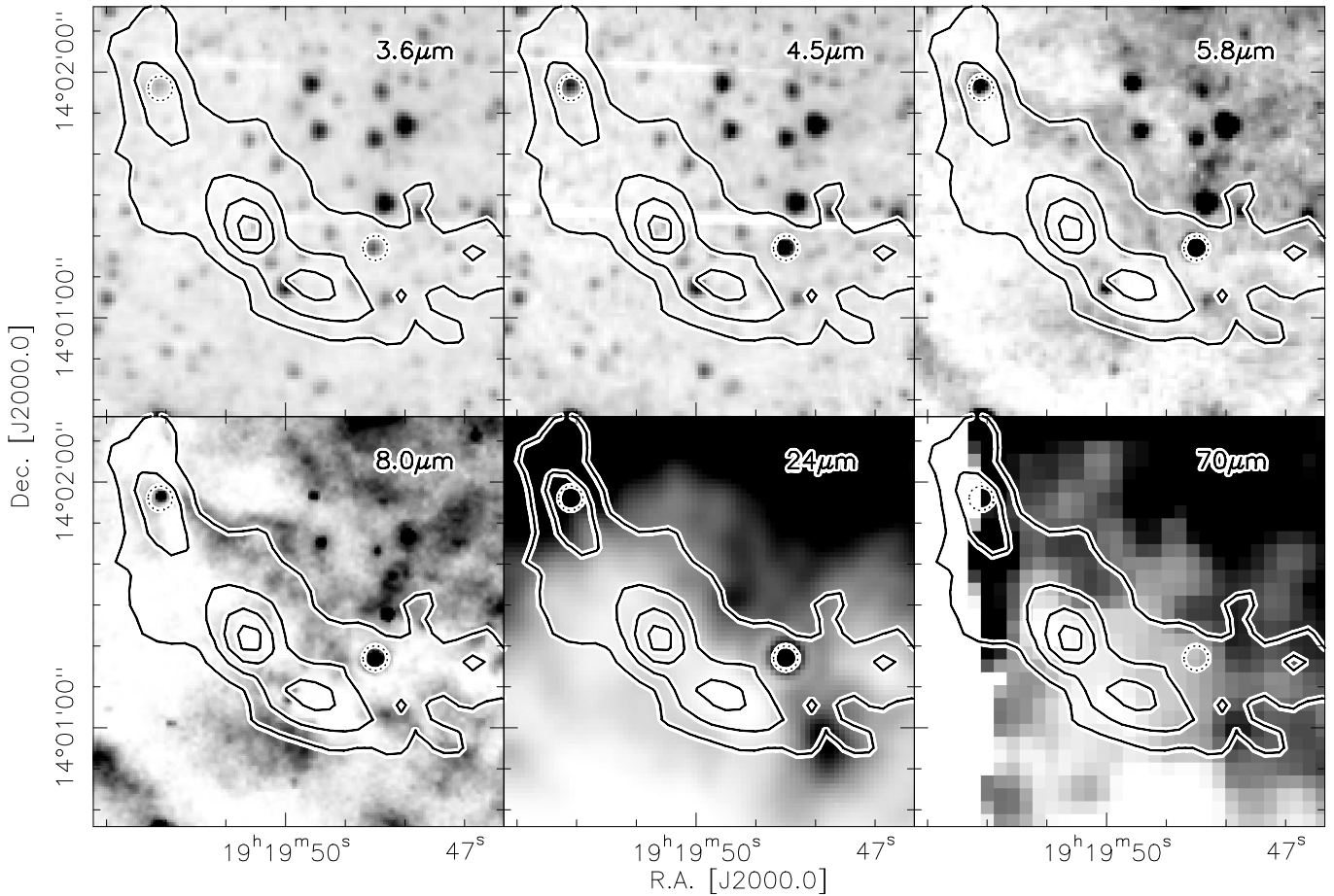}
\caption{Zoom into the central region of the two target sources.  Like
  Figure \ref{overview}, the contours show the 1.2\,mm dust continuum
  emission \citep{beuther2002a}, and the grey-scale presents the mid-
  to far-infrared emission as observed with the Spitzer Space
  Telescope.  From top-left to bottom-right, the 6 panels show the
  Spitzer bands centered at 3.6, 4.5, 5.8, 8.0, 24 and 70\,$\mu$m, as
  marked in each panel. The dotted circles in each panel mark the only
  two Spitzer sources that were identified as class0/I candidates (see
  main text).}
\label{spitzer}
\end{figure*}

Figure \ref{spitzer} presents an overlay of all IRAC and the MIPS 24
and 70\,$\mu$m bands \citep{churchwell2009,carey2009} with the dust
continuum emission towards IRDC\,19175-4 and IRDC\,19175-5, and both
regions remain dark even up to 70\,$\mu$m. The 24\,$\mu$m absorption
feature toward IRDC\,19175-5 is a bit more pronounced than toward
IRDC\,19175-4. We detected no SiO emission. Their NH$_3$ single-dish
line widths are $\sim$1.5\,km/s, and their derived NH$_3$ rotation
temperatures are of the order 15\,K \citep{sridha,beuther2007g}. All
these characteristics make the two sources ideal HMSC candidates and
hence well-suited targets to investigate the physical properties prior
to the onset of star formation.

Assuming optically thin emission from cold dust at 15\,K and a
distance of 1.1\,kpc with a gas-to-dust mass ratio of 186 (following
\citealt{draine2007,jenkins2004}), the integrated 1.2\,mm emission
from the two region IRDC\,19175-4 and IRDC\,19175-5 of 620\,mJy
\citep{beuther2002a} corresponds to approximately 87\,M$_{\odot}$ of
gas mass, fairly equally divided between both sources.  While these
gas masses are not of the order 1000\,M$_{\odot}$ like some better
known massive star-forming regions, the gas reservoir is large enough
to potentially form intermediate-mass B-type stars.

\section{Observations}

We observed IRDC\,19175-4 and IRDC19175-5 with the Plateau de Bure
Interferometer during two nights in August 2007 and April 2008 at
93\,GHz in the C and D configurations covering projected baselines
between approximately 13 and 180\,m.  The 3\,mm receivers were tuned
to 92.835\,GHz in the lower sideband covering the N$_2$H$^+$(1--0) and
the $^{13}$CS(2--1) lines as well as the 3.23\,mm continuum emission.
The phase noise was always lower than 30$^{\circ}$ and mostly lower
than 20$^{\circ}$. For continuum measurements we placed six 320\,MHz
correlator units in the band, the spectral lines were excluded in
averaging the units to produce the final continuum image.  Temporal
fluctuations of amplitude and phase were calibrated with frequent
observations of the quasars 1923-201 and 1827+062.  The amplitude
scale was derived from measurements of MWC349. We estimate the final
flux accuracy to be correct to within $\sim 15\%$. The phase reference
center is R.A.[J2000] 19$^h$19$^m$50.169$^s$ and Dec.[J2000]
14$^{\circ}$01$'$13.75$''$, and the velocity of rest $v_{\rm{lsr}}$ is
7.7\,km\,$s^{-1}$. The synthesized beam of the continuum data with a
robust weighting is $3.5''\times 2.8''$ (P.A.  $28^{\circ}$ east of
north) with a $3\sigma$ continuum rms of 0.16\,mJy\,beam$^{-1}$. The
$3\sigma$ rms of the N$_2$H$^+$(1--0) data measured from an
emission-free channel with a spectral resolution of 0.2\,km\,s$^{-1}$
is 15\,mJy\,beam$^{-1}$. The $3\sigma$ rms of the $^{13}$CS(2--1) data
measured from an emission-free channel with a spectral resolution of
0.5\,km\,s$^{-1}$ is 10\,mJy\,beam$^{-1}$.

These mm observations are complemented with the mid-infrared SPITZER
data from the GLIMPSE and MIPSGAL surveys of the Galactic plane using
the IRAC camera centered at 3.6, 4.5, 5.8 and 8.0\,$\mu$m and the MIPS
camera centered at 24 and 70\,$\mu$m
\citep{werner2004,fazio2004,benjamin2003,rieke2004,carey2005}.

\section{Results}

\begin{figure}[htb] 
\centering
\includegraphics[angle=-90,width=0.48\textwidth]{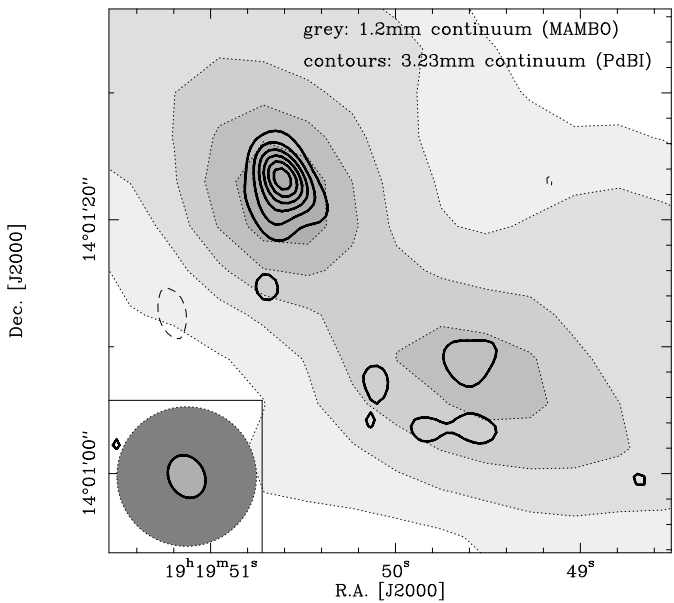}
\caption{Millimeter continuum emission toward IRDC\,19175-4 and
  IRDC19175-5. The grey-scale with dotted contours presents the
  single-dish 1.2\,mm continuum observations with the MAMBO array
  mounted on the IRAM 30\,m telescope contoured at the $3\sigma$
  levels of 15\,mJy\,beam$^{-1}$ (Fig.\ref{overview} and
  \citealt{beuther2002a}), and the thick contours show the new
  3.23\,mm data from the PdBI observed for this project with $3\sigma$
  contours of 0.16\,mJy\,beam$^{-1}$. Full and dashed contours present
  emission and negative features due to the insufficient uv-coverage,
  respectively. In the bottom-left corner, the beam of the 1.2\,mm
  single-dish observations is shown as the large circle, and the
  synthesized PdBI beam as smaller ellipse.}
\label{continuum} 
\end{figure}

\subsection{Millimeter continuum emission and class 0/I sources
  identified via the Spitzer data}
\label{cont}

Figure \ref{continuum} presents an overlay of the 1.2\,mm dust
continuum data from the IRAM 30\,m telescope \citep{beuther2002a} with
the new high-spatial-resolution 3.23\,mm data obtained now with the
PdBI.  While the north-eastern peak (IRDC\,19175-4) is clearly
detected also in the new 3.23\,mm data, the south-western peak
(IRDC\,19175-5) is a marginal detection at less than $5\sigma$ with
0.26\,mJy\,beam$^{-1}$.  The peak and integrated 3.23\,mm fluxes of
IRDC\,19175-4 are 1.09\,mJy\,beam$^{-1}$ and 1.8\,mJy, respectively.
Following \citet{hildebrand1983} and \citet{beuther2002a}, assuming
optically thin dust continuum emission with a gas-to-dust mass ratio
of 186 \citep{draine2007,jenkins2004}, a dust opacity index $\beta =
2$ (corresponding to a typical dust distribution as outlined in
\citealt{mathis1977,ossenkopf1994}) and a temperature of 15\,K
\citep{sridharan2005}, we derive a gas mass and column density of
$\sim$10\,M$_{\odot}$ and $\sim$1.8$\times 10^{24}$\,cm$^{-2}$,
respectively. The latter corresponds to a visual extinction $A_v$ of
$\sim$1900\,mag \citep{frerking1982}. Our $3\sigma$ continuum
sensitivity corresponds with the given parameters to $3\sigma$ mass
and column density sensitivities of 0.9\,M$_{\odot}$ and 1.7$\times
10^{23}$\,cm$^{-2}$.  Comparing the mass derived from the PdBI data of
IRDC\,19175-4 of $\sim$10\,M$_{\odot}$ with the single-dish estimated
mass of 87\,M$_{\odot}$, approximately 90\% of the flux has been
filtered out by the interferometer observations.

Although the mm continuum peaks are infrared dark up to 70\,$\mu$m
emission, Figure \ref{spitzer} shows several Spitzer sources within
the field of view. Employing the available IRAC GLIMPSE point source
catalog (\citealt{churchwell2009} \& http://irsa.ipac.caltech.edu/)
and adopting the Spitzer color criteria by \citet{allen2004}, we tried
to identify possible class 0/I and class II protostellar candidates
around the discussed IRDCs (see also Fang et al., subm.).
Interestingly, we do not find any class II source and only 2 class 0/I
candidates close to IRDC\,19175-4 and IRDC\,19175-5 (see dotted
circles in Fig.~\ref{spitzer}). While these two class 0/I candidates
are clearly within the vicinity of the IRDCs, almost all of the
remaining sources in the field of view are either more evolved and
hence do not fit the Spitzer color criteria, or they are entirely
unrelated fore- and/or background sources. Identifying only two young
protostellar candidates with locations not toward the mm continuum
peaks further stresses the youth of the target regions.

\begin{figure}[htb]
  \centering
  \includegraphics[angle=-90,width=0.48\textwidth]{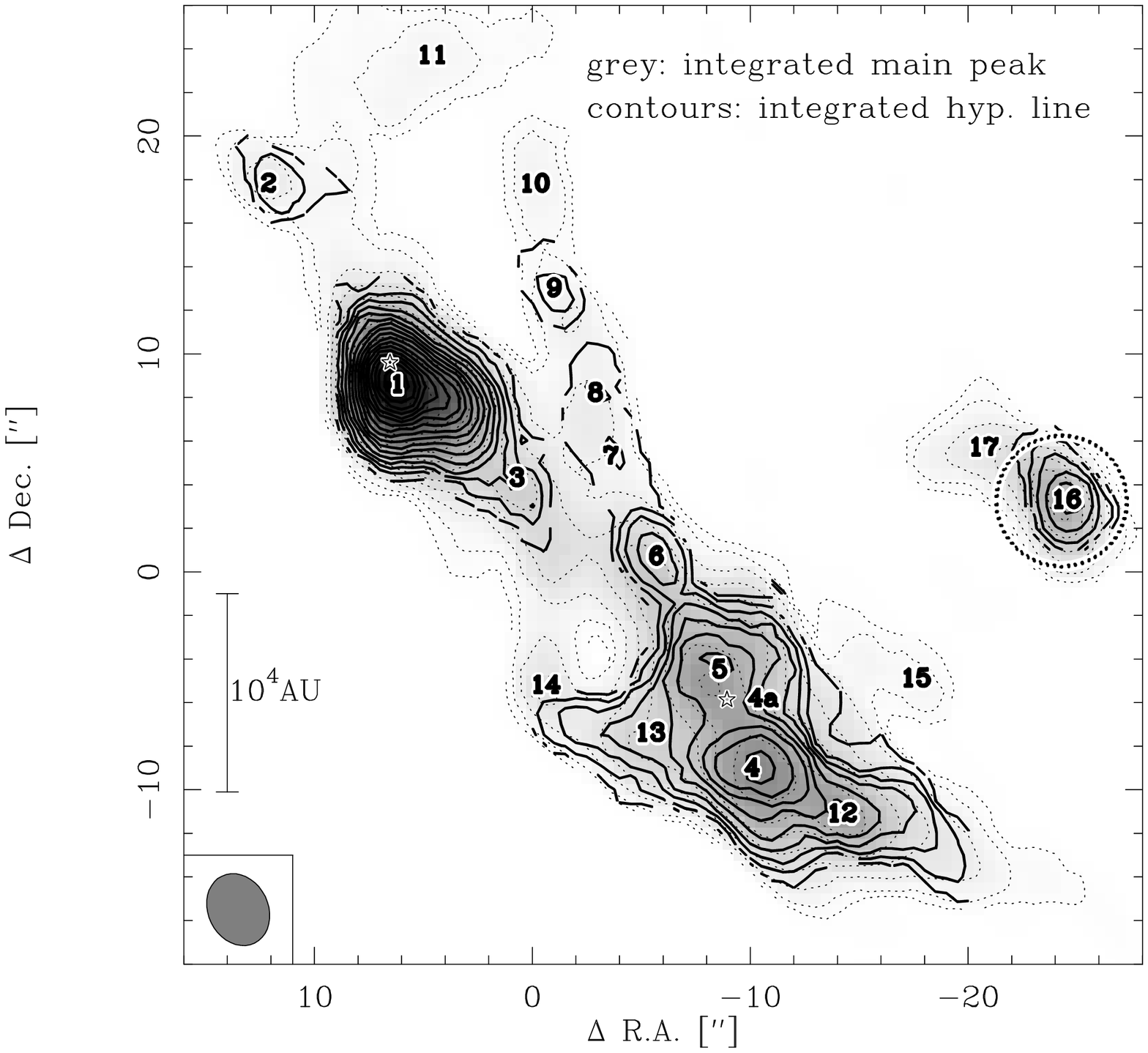}
  \caption{The thick contours present the integrated emission of the
    isolated N$_2$H$^+$(1--0) (F1, F=0,1$\rightarrow$1,2) hyperfine
    emission component at 93176.2650\,MHz \citep{caselli1995} which is
    blue-shifted by -8.0\,km\,s$^{-1}$ with respect to the central
    component (see Fig.~\ref{spectra}). The contour levels are in
    steps of 10\,mJy\,beam$^{-1}$\,km\,s$^{-1}$.  The grey-scale with
    dotted contours shows the integrated emission of the central few
    hyperfine components around the $v_{\rm{lsr}}$ with contour levels
    of 20, 40 and then continuing in
    40\,mJy\,beam$^{-1}$\,km\,s$^{-1}$ steps. To produce the
    integrated intensity (or 0th moment) maps, data were clipped below
    $5\sigma$ corresponding to 25\,mJy\,beam$^{-1}$ for a channel
    width of 0.2\,km\,s$^{-1}$. The corresponding integration regimes
    are 6.3 to 8.9\,km\,s$^{-1}$ for the isolated hyperfine structure
    line and 5.5 to 10.7\,km\,s$^{-1}$ for the blended central line,
    respectively.  The stars mark the positions of the main mm
    continuum peaks (Figs.~\ref{overview} \& \ref{continuum}), and the
    synthesized beam and scale-bar are shown at the bottom-left.  The
    numbers mark the 17 N$_2$H$^+$(1--0) emission peaks discussed in
    the main text and Table \ref{n2h+}.  The dotted circle associated
    with the N$_2$H$^+$ peak 16 marks the position of the western
    Spitzer class 0/I source. The offsets are given with respect to
    the phase center.}
\label{overlay} 
\end{figure}

\begin{figure*}[htb]
  \centering
  \includegraphics[angle=-90,width=0.98\textwidth]{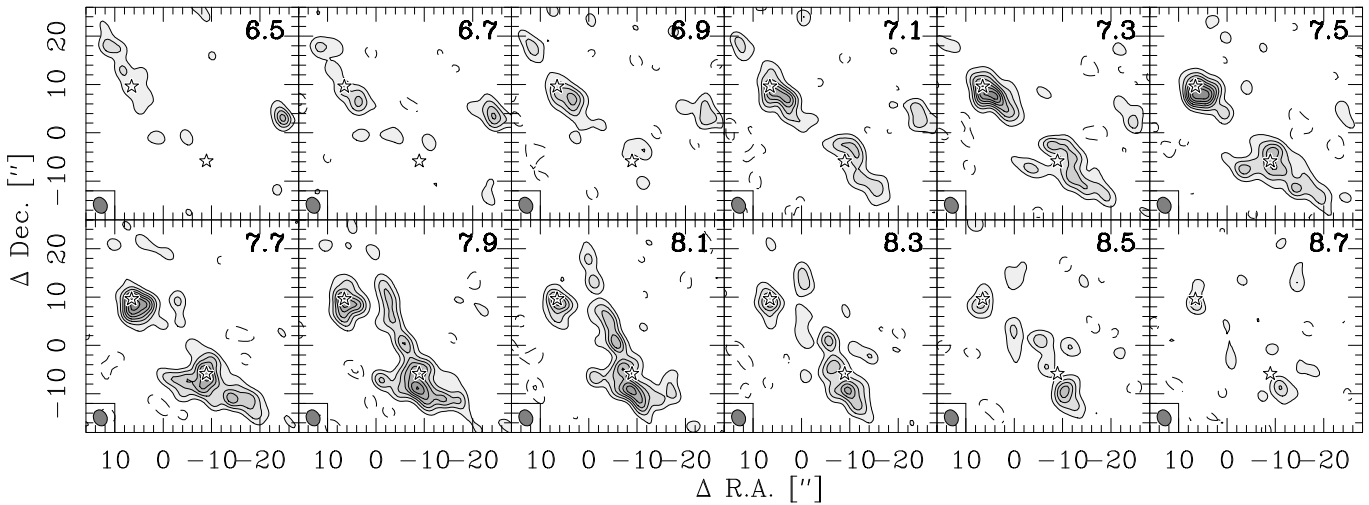}
  \caption{Channel map of the isolated N$_2$H$^+$(1--0) (F1,
    F=0,1$\rightarrow$1,2) hyperfine emission component at
    93176.2650\,MHz \citep{caselli1995} which is blue-shifted by
    -8.0\,km\,s$^{-1}$ with respect to the central component (see
    Fig.~\ref{spectra}). The velocities as marked in the top-right
    corners of each panel are shifted to the velocity of rest. The
    grey-scale and contours are done in $3\sigma$ steps of
    15\,mJy\,beam$^{-1}$ (full and dashed contours positive and
    negative, respectively). The two stars mark the main two mm
    continuum sources (Figs.~\ref{overview} \& \ref{continuum}), the
    numbers in the top-right corners show the central velocities of
    each channel, and the synthesized beam is plotted at the
    bottom-left of each panel. The offsets are given with respect to
    the phase center.}
\label{channel} 
\end{figure*}

\subsection{N$_2$H$^+$(1--0) emission} 
\label{n2h+_main}

\begin{figure*}[htb] 
\centering
  \includegraphics[angle=-90,width=\textwidth]{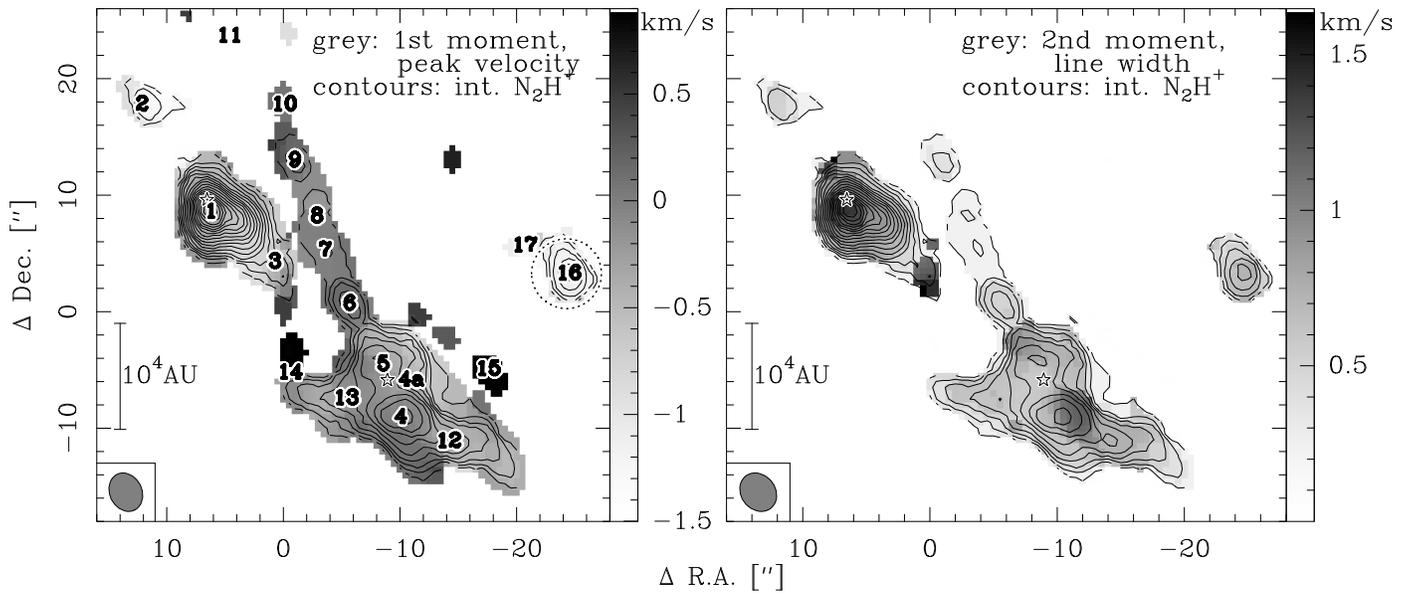}
  \caption{Moment maps of the N$_2$H$^+$(1--0) emission. The
    grey-scale in the left and right panels present the 1st and 2nd
    moments, respectively (intensity-weighted peak velocities and line
    widths).  The contours in both panels show the 0th moment or
    integrated intensity maps which were produced from the channel map
    in Fig.~\ref{channel} via clipping all data below $5\sigma$
    corresponding to 25\,mJy\,beam$^{-1}$ for a channel width of
    0.2\,km\,s$^{-1}$. The contour levels are done in steps of
    10\,mJy\,beam$^{-1}$\,km\,s$^{-1}$. The maps were produced from
    the isolated N$_2$H$^+$(1--0) (F1, F=0,1$\rightarrow$1,2)
    hyperfine emission component (-8.0\,km\,s$^{-1}$ offset from the
    central component) with an integration regime corresponding to 6.3
    to 8.9\,km\,s$^{-1}$ ($v_{\rm{lsr}}=7.7$\,km\,$s^{-1}$). The stars
    mark the positions of the main mm continuum peaks
    (Figs.~\ref{overview} \& \ref{continuum}), and the synthesized
    beam and scale-bar are shown at the bottom-left of each panel. The
    numbers in the left panel mark the 17 N$_2$H$^+$(1--0) emission
    peaks discussed in the main text and Table \ref{n2h+}. The dotted
    circle associated with the N$_2$H$^+$ peak 16 marks the position
    of the western Spitzer class 0/I source. The offsets are given
    with respect to the phase center.}
\label{moments} 
\end{figure*}

\begin{figure}[htb] 
\centering
\includegraphics[angle=-90,width=0.48\textwidth]{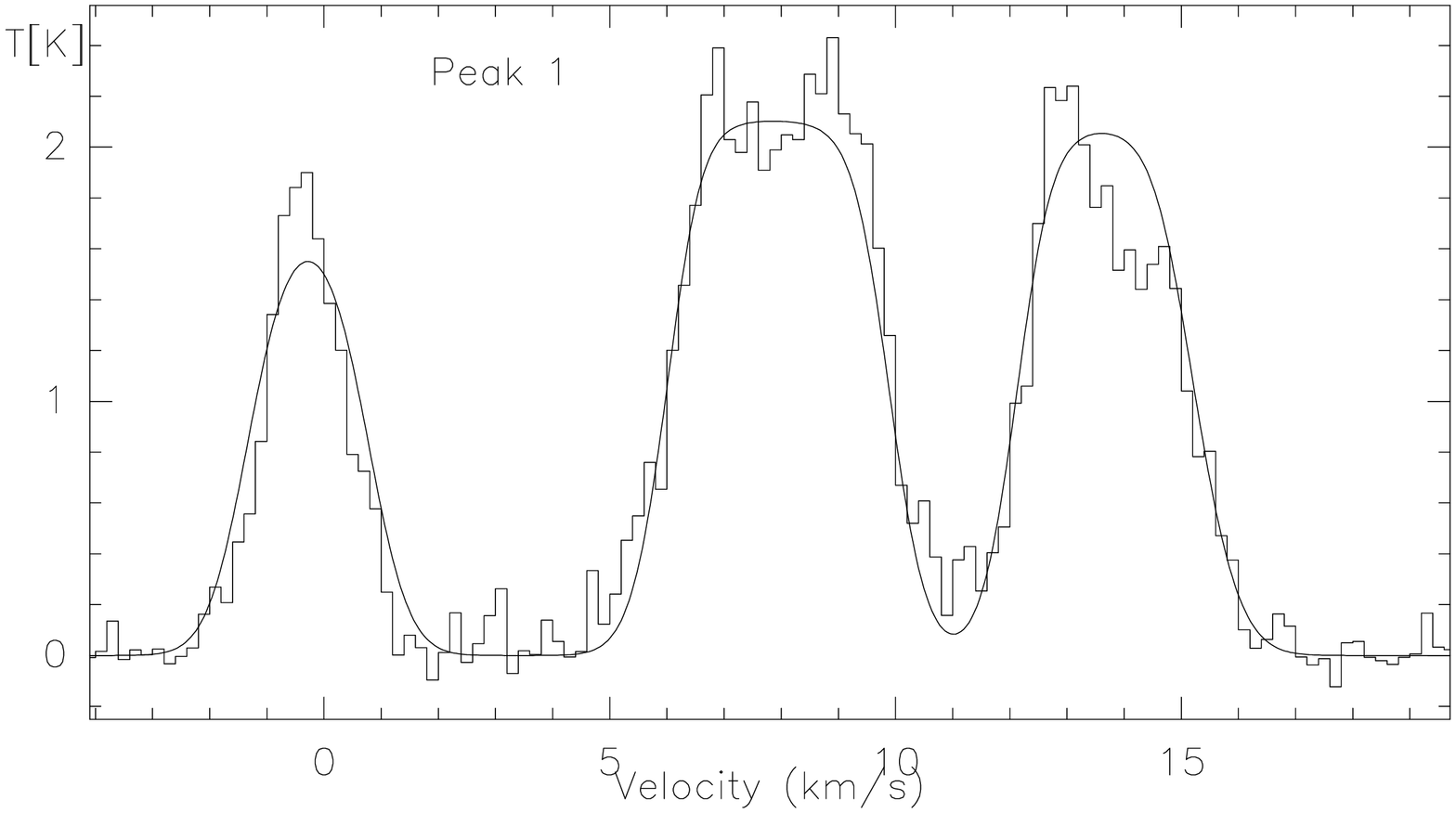}\\
\includegraphics[angle=-90,width=0.48\textwidth]{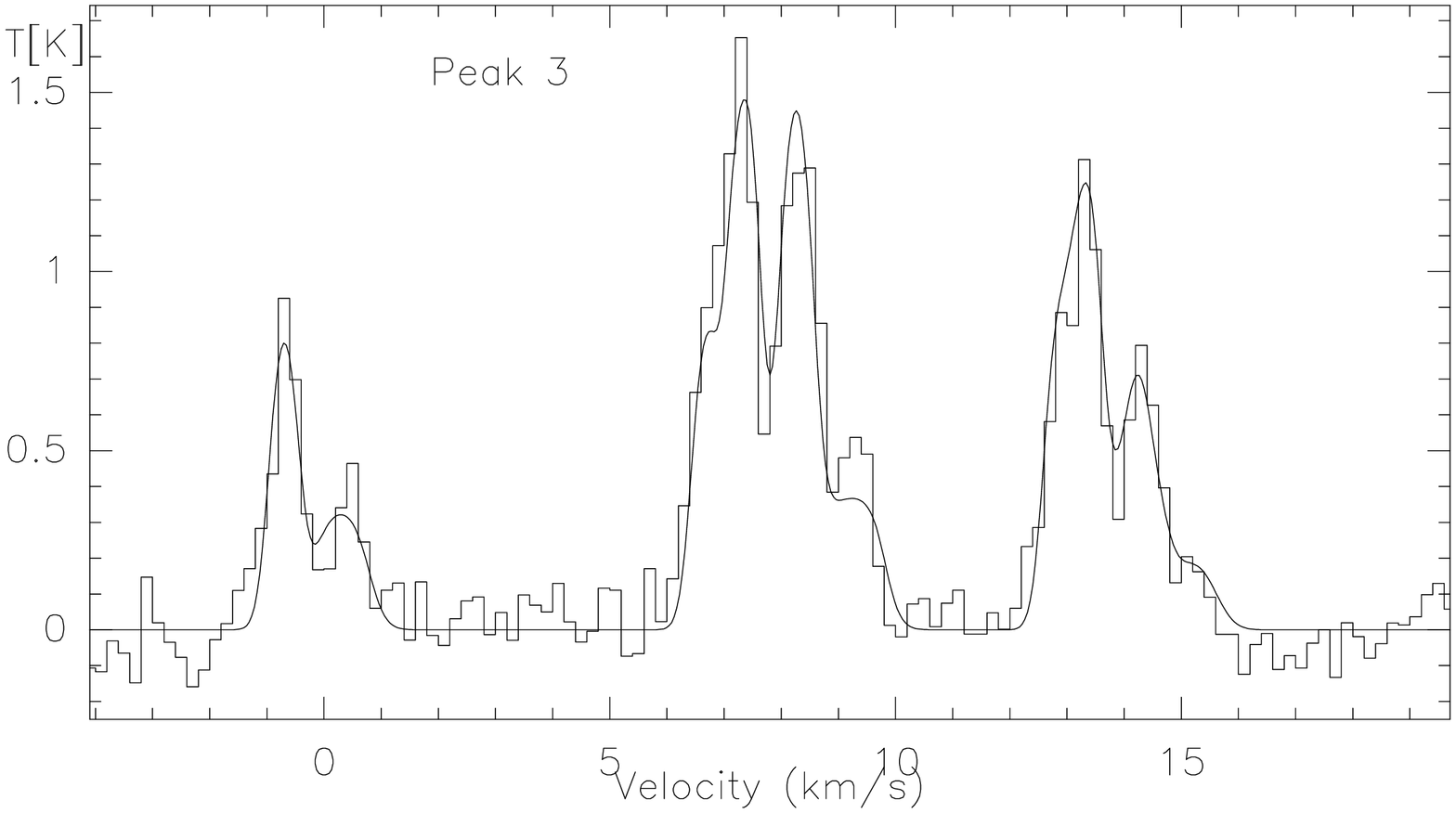}\\
\includegraphics[angle=-90,width=0.48\textwidth]{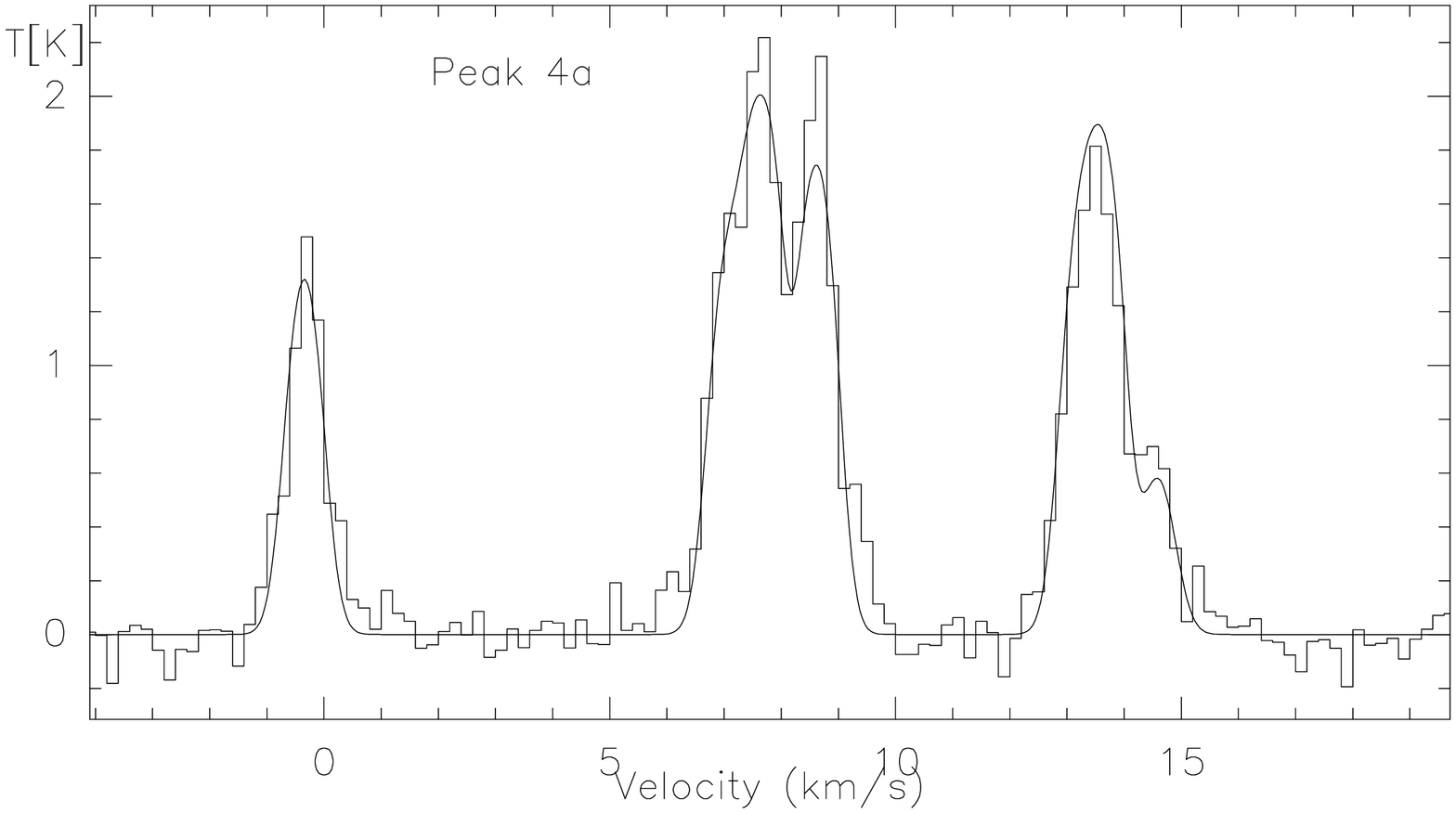}\\
\includegraphics[angle=-90,width=0.48\textwidth]{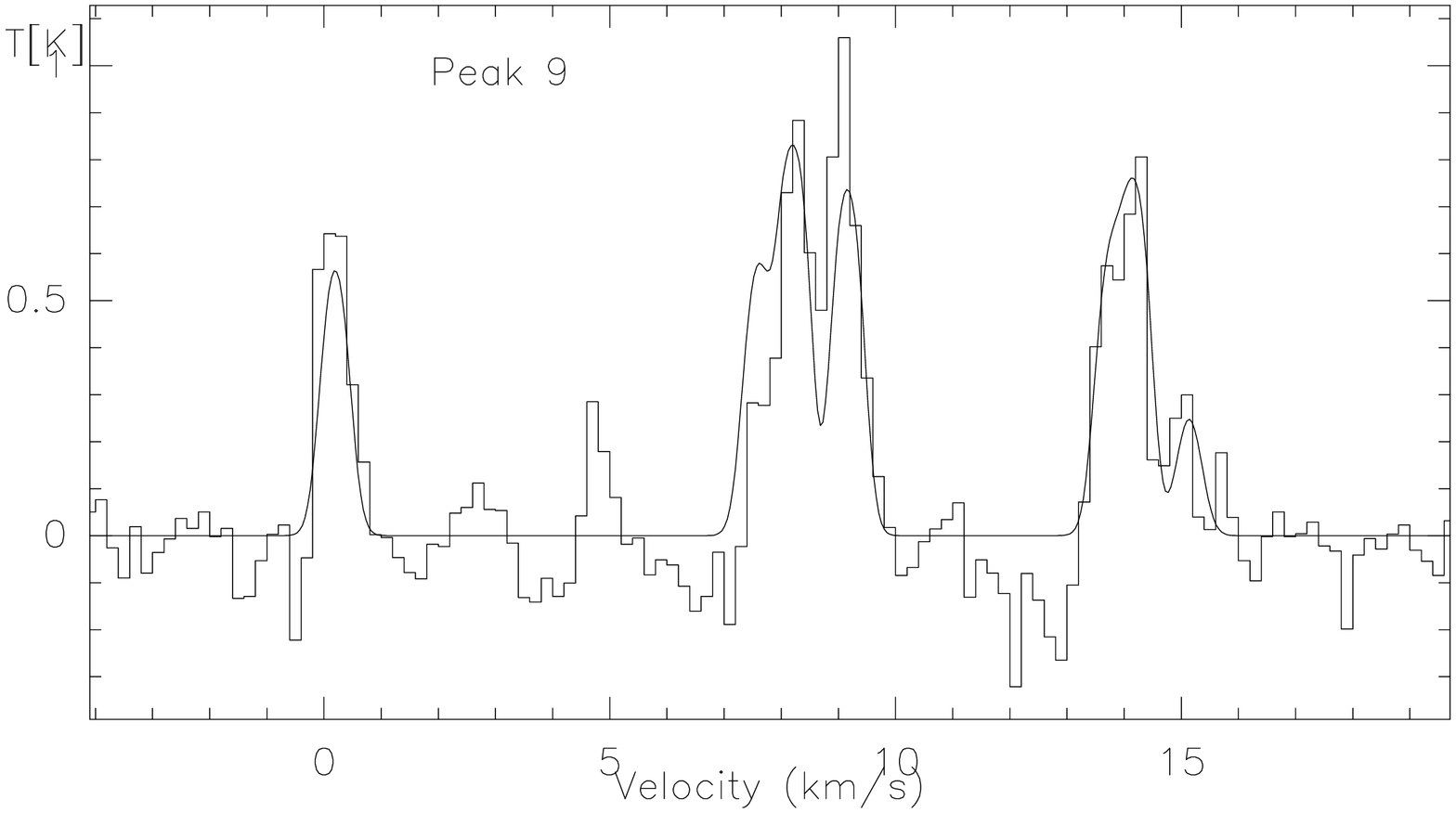}
\caption{Example N$_2$H$^+$(1--0) spectra extracted toward the
  positions labeled in each panel (see Figure \ref{moments} and Table
  \ref{n2h+}). The full lines show the fits with the parameters listed
  in Table \ref{n2h+}. The y-axis shows the brightness temperature.}
\label{spectra}
\end{figure}

Figures \ref{overlay}, \ref{channel} and \ref{moments} present the
N$_2$H$^+$(1--0) data as integrated intensity maps, a channel map and
the deduced 1st and 2nd moment maps (intensity-weighted peak
velocities and line widths). For the channel and moment maps, we use
the isolated N$_2$H$^+$(1--0) (F1, F=0,1$\rightarrow$1,2) hyperfine
emission component at 93176.2650\,MHz \citep{caselli1995} which is
blue-shifted by -8.0\,km\,s$^{-1}$ with respect to the central
component because it is a spectrally isolated line without further
line-blending and mostly optically thin. In strong contrast to the
relatively simple dust continuum structure, the N$_2$H$^+$ data now
reveal a large amount of separate emission peaks, 17 in number.
Although the general structure of the two integrated maps of the main
and satellite hyperfine structure lines in Figure \ref{overlay} is
very similar, small discrepancies arise because of different
sensitivities and opacities. While N$_2$H$^+$ peaks 10, 11, 14, 15 and
17 are mainly detected in the main hyperfine structure line most
likely because it can trace lower column density structures, other
peaks are offset between the two integrated images. For example the
N$_2$H$^+$ peaks 7 and 8 are clearly separated in the satellite
hyperfine structure line map, whereas it appears like a single peak in
the main hyperfine line.  This difference likely arises from the
higher optical depth of the main line. For our source-labeling we
refer to the positions from the isolated satellite hyperfine structure
line where they were clearly detected. For the remaining peaks, the
positions are extracted from the main hyperfine line.  While the
strongest N$_2$H$^+$ peak number 1 is associated with the strongest mm
continuum peak, the second mm peak has no directly associated
N$_2$H$^+$ feature, rather the N$_2$H$^+$ peaks 4 and 5 are lying at
the edge of the mm continuum emission.  Figure \ref{spectra} presents
a few example N$_2$H$^+$ spectra, and Table \ref{n2h+} shows the
corresponding spectral parameters (peak line temperature
$T_{\rm{peak}}$, peak velocity $v$, line width $\Delta v$ and total
optical depth $\tau$ of all hyperfine components) we derived from
fitting the complete N$_2$H$^+$(1--0) hyperfine structure within CLASS
(part of the GILDAS software
package\footnote{http://www.iram.fr/IRAMFR/GILDAS}).  While the
N$_2$H$^+$ peak 16 is spatially associated with the western Spitzer
class 0/I source (Figs.~\ref{spitzer} \& \ref{moments}), no other
N$_2$H$^+$ peak has a Spitzer-identifiable young protostellar
counterpart. Except of peak 16, all N$_2$H$^+$ peaks are good
candidates for being genuine starless cores.

\begin{table*}[htb] 
\centering 
\caption{Parameters of N$_2$H$^+$(1--0) emission peaks} 
\begin{tabular}{lrrrrrrrr} \hline \hline
  \# & Offset$^a$ & $T_{\rm{peak}}$ & $v$ & $\Delta v$ & $\tau$ & $N_{\rm{N_2H^+}}$ & $M_{\rm{vir}}^b$ & av.~$\rho$ \\
  & ''     & K               & km/s& km/s       &        & $\times 10^{13}$cm$^{-2}$ & M$_{\odot}$ & $10^5$cm$^{-3}$ \\
  \hline
  1  & 6.2/8.6     & 2.1 & 7.7 & 1.8 & 11.9 & $6.6\pm 0.3$ & 7.5-11.4 & 19\\
  2  & 12.1/17.8   & 0.7 & 6.9 & 0.8 & 12.7 & $1.9\pm 0.3$ & 0.6-0.9  & 12\\
  $3_1$& 0.7/4.3   & 1.5 & 7.3 & 0.5 &  7.0 & $0.9\pm 0.1$ & 0.3-0.4  & 5.9\\
  $3_2$& 0.7/4.3   & 0.4 & 8.3 & 0.8 & 16.3 & $2.1\pm 0.5$ & 0.6-0.9  & 14\\
  4  & -10.1/-9.0  & 2.3 & 8.1 & 1.1 &  6.5 & $2.3\pm 0.2$ & 1.3-1.9  & 15\\
  4a & -9.9/-5.8   & 2.3 & 7.7 & 0.7 &  7.8 & $1.7\pm 0.1$ & 0.5-0.7  & 11\\
  5  & -8.6/-4.5   & 2.2 & 7.7 & 0.8 &  7.1 & $1.9\pm 0.1$ & 0.7-1.1  & 12\\
  6  & -5.7/0.7    & 1.3 & 8.2 & 0.7 & 12.6 & $2.0\pm 0.1$ & 0.5-0.7  & 13\\
  7  & -3.6/5.3    & 1.3 & 8.1 & 0.6 &  5.1 & $0.7\pm 0.1$ & 0.3-0.5  & 4.6\\
  8  & -2.9/8.2    & 4.3 & 8.0 & 0.7 &  0.1 & $0.8\pm 0.1$ & 0.5-0.7  & 5.2\\
  9  & -1.0/13.0   & 0.9 & 8.2 & 0.5 &  8.3 & $0.9\pm 0.2$ & 0.3-0.4  & 5.9 \\
  10 & -0.1/17.8   & 2.2 & 8.1 & 0.5 &  2.0 & $0.4\pm 0.1$ & 0.3-0.5  & 2.6 \\
  $11_1$ & 4.6/23.7& 1.4 & 7.6 & 0.5 &  1.1 & $0.3\pm 0.1$ & 0.3-0.5  & 2.0 \\
  $11_2$ & 4.6/23.7& 0.6 & 8.6 & 0.8 &  4.9 & $0.7\pm 0.1$ & 0.6-0.9  & 4.6\\
  12 & -14.2/-11.1 & 2.0 & 7.7 & 1.2 &  5.7 & $2.0\pm 0.2$ & 1.5-2.2  & 13 \\
  13 & -5.4/-7.4   & 1.5 & 7.8 & 0.7 & 10.4 & $1.8\pm 0.1$ & 0.5-0.7  & 12\\
  14$_1$& -0.6/-5.2   & 0.5 & 7.8 & 1.0 &  10.7 & $1.9\pm 0.2$ & 1.0-1.5  & 12\\
  14$_2$& -0.6/-5.2   & 2.6 & 8.8 & 0.7 &  0.1 & $0.3\pm 0.1$ & 0.5-0.7  & 2.0\\
  15 & -17.6/-4.9  & 2.6 & 8.9 & 0.5 &  0.9 & $0.3\pm 0.3$ & 0.3-0.4  & 2.0\\
  16 & -24.5/3.3   & 2.8 & 6.8 & 1.0 &  3.7 & $1.4\pm 0.1$ & 0.9-1.4  & 9.2\\
  17 & -20.7/5.7   & 3.1 & 6.8 & 1.0 &  0.1 & $0.6\pm 0.2$ & 1.0-1.6  & 3.9\\
  \hline \hline \end{tabular}
~\\
{\footnotesize 
  The subscripts associated with sources 3, 11 and 14 correspond to two different velocity components required for the spectral fits.\\
  $^a$ The offsets are given with respect to the phase center  R.A.[J2000] 
  19$^h$19$^m$50.169$^s$ and Dec.[J2000] 14$^{\circ}$01$'$13.75$''$.\\
  $^b$ The virial masses were calculated following
  \citet{maclaren1988} for $1/r$ and $1/r^2$ density distributions,
  resulting in the ranges of values. All source radii were
  approximated with half of the average N$_2$H$^+$ synthesized beam
  of $\sim 3.1''$. Only for peak one we used a source radius of
  $3.5''$.}  
\label{n2h+} 
\end{table*}

The sub-structure of the cores exhibits an elongated and filamentary
structure. The filament associated with the N$_2$H$^+$ peaks 6 to 10
has a linear extend of $\sim$25000\,AU with a length-to-width ratio
exceeding 5. The average separation between the five cores is
$\sim$5000\,AU. Following \citet{stahler2005} we can calculate the
Jeans-length

$$\lambda_J = \left(\frac{\pi a^2}{G\rho}\right)^{1/2} = 0.19\rm{pc}\left(\frac{T}{10\rm{K}}\right)^{1/2}\left(\frac{n_{\rm{H_2}}}{10^4\rm{cm}^{-3}}\right)^{-1/2}$$

with the sound speed $a$, the gravitational constant $G$ and the
density $\rho$. What density do we expect for the cores? The critical
density of the N$_2$H$^+$(1--0) transition at the given low
temperature is $\sim 1.5\times 10^5$\,cm$^{-3}$. In addition to this,
we can calculate average densities for each core converting the
N$_2$H$^+$ column densities (see section \ref{column}) to H$_2$ column
densities assuming a typical N$_2$H$^+$ to H$_2$ abundance ratio of
$3\times 10^{-10}$ \citep{caselli2002b}, and furthermore assuming in
spherical geometry that the 3rd dimension along the line of sight
corresponds to the synthesized beam for the compact cores.  Table
\ref{n2h+} lists the derived densities for all cores estimated this
way. The range of average densities varies by about one order of
magnitude between $2\times 10^5$ and $1.9\times 10^6$\,cm$^{-3}$. For
the cores 6 to 10 along the filament, the average density is about
$\sim 6\times 10^5$\,cm$^{-3}$, implying that the peak densities
should be even higher. Hence, we can estimate the Jeans-length at
densities of $5\times 10^5$ and $10^6$\,cm$^{-3}$ with a typical
temperature of 15\,K. This brackets the corresponding Jeans-lengths
approximately between $\sim$4800 and $\sim$6800\,AU above which the
filament should become unstable.  Interestingly, the measured core
separation corresponds well to this Jeans-length indicating that the
cores could be stable against further fragmentation.

While the broadest line (still only 1.8\,km\,s$^{-1}$ full width half
maximum) is associated with the main mm continuum peak, all other peak
positions exhibit line widths close to or below 1\,km\,s$^{-1}$. The
N$_2$H$^+$(1--0) thermal line width at 15\,K is
$\sim$0.15\,km\,s$^{-1}$. Since the spectral resolution of our final
N$_2$H$^+$(1--0) data cube (0.2\,km\,s$^{-1}$) is insufficient to
actually resolve the thermal line width, it is difficult to accurately
constrain whether the narrowest measured line widths (Table
\ref{n2h+}) close to our spectral resolution (less than 3 spectral
resolution elements across the line width $\Delta v$) are not rather
upper limits to real line widths closer to thermal. 

The relative peak velocities between close-by neighboring sub-cores is
usually below 1\,km\,s$^{-1}$, and we do not identify any clear
velocity gradient or streaming motions along the filamentary
structures associated with the N$_2$H$^+$ peak positions (see Figure
\ref{moments} and Table\ref{n2h+}).

\subsection{N$_2$H$^+$ column densities and abundances}
\label{column}

Using the line intensities and optical depths from our
N$_2$H$^+$(1--0) fitting, we can estimate the N$_2$H$^+$ column
densities for all cores (e.g, \citealt{caselli2002b}).  The resulting
values between $0.3\times 10^{13}$ and $6.6\times 10^{13}$\,cm$^{-2}$
are also listed in Table \ref{n2h+}. With a typical N$_2$H$^+$/H$_2$
abundance ratio of $3\times 10^{-10}$ \citep{caselli2002b}, this
corresponds to an H$_2$ column density regime between roughly $2\times
10^{22}$ and $2\times 10^{23}$\,cm$^{-1}$ (or visual extinctions $A_v$
between 20 and 200, \citealt{frerking1982}), largely below our H$_2$
$3\sigma$ column density limit of $1.7\times 10^{23}$\,cm$^{-2}$
derived from the 3.23\,mm dust continuum emission (see section
\ref{cont}). However, taking our derived H$_2$ and N$_2$H$^+$ column
densities at face value, we derive an order of magnitude lower
abundance of $\sim 3.7\times 10^{-11}$ for the main 3.23\,mm continuum
and N$_2$H$^+$ peak 1 associated with IRDC\,19175-4. Assuming that
this abundance could hold for all N$_2$H$^+$(1--0) cores, the range of
corresponding H$_2$ column densities would be an order of magnitude
higher, roughly between $2\times 10^{23}$ and $2\times
10^{24}$\,cm$^{-2}$. These values nearly all were above our H$_2$
$3\sigma$ column density limit of $1.7\times 10^{23}$\,cm$^{-2}$
derived from the 3.23\,mm continuum emission. Although all these
estimates are prone to many uncertainties, they indicate that even
during such early evolutionary stages the gas phase abundances are not
uniform throughout the core.  The source with the broadest line width,
and hence likely the most evolved of the cores (peak 1 in
IRDC\,19175-4), shows the lowest abundance compared to the other less
turbulent cores.  While \citet{tafalla2004b} argued in the case of the
low-mass starless core L1571E that particular low N$_2$H$^+$
abundances could also be due to very young evolutionary stages (see
also \citealt{aikawa2003}), the increased line width toward our main
core makes this an unlikely scenario. Rather in contrast, increased
densities in more evolved stages of core collapse (like those inferred
here, see Table \ref{n2h+}) increase the recombination rate and hence
lower the N$_2$H$^+$ abundance (e.g, \citealt{bergin1997}). This
appears the more likely scenario for the cores we observe here.

\subsection{$^{13}$CS(2--1) emission}

\begin{figure}[htb] 
\centering
\includegraphics[angle=-90,width=0.48\textwidth]{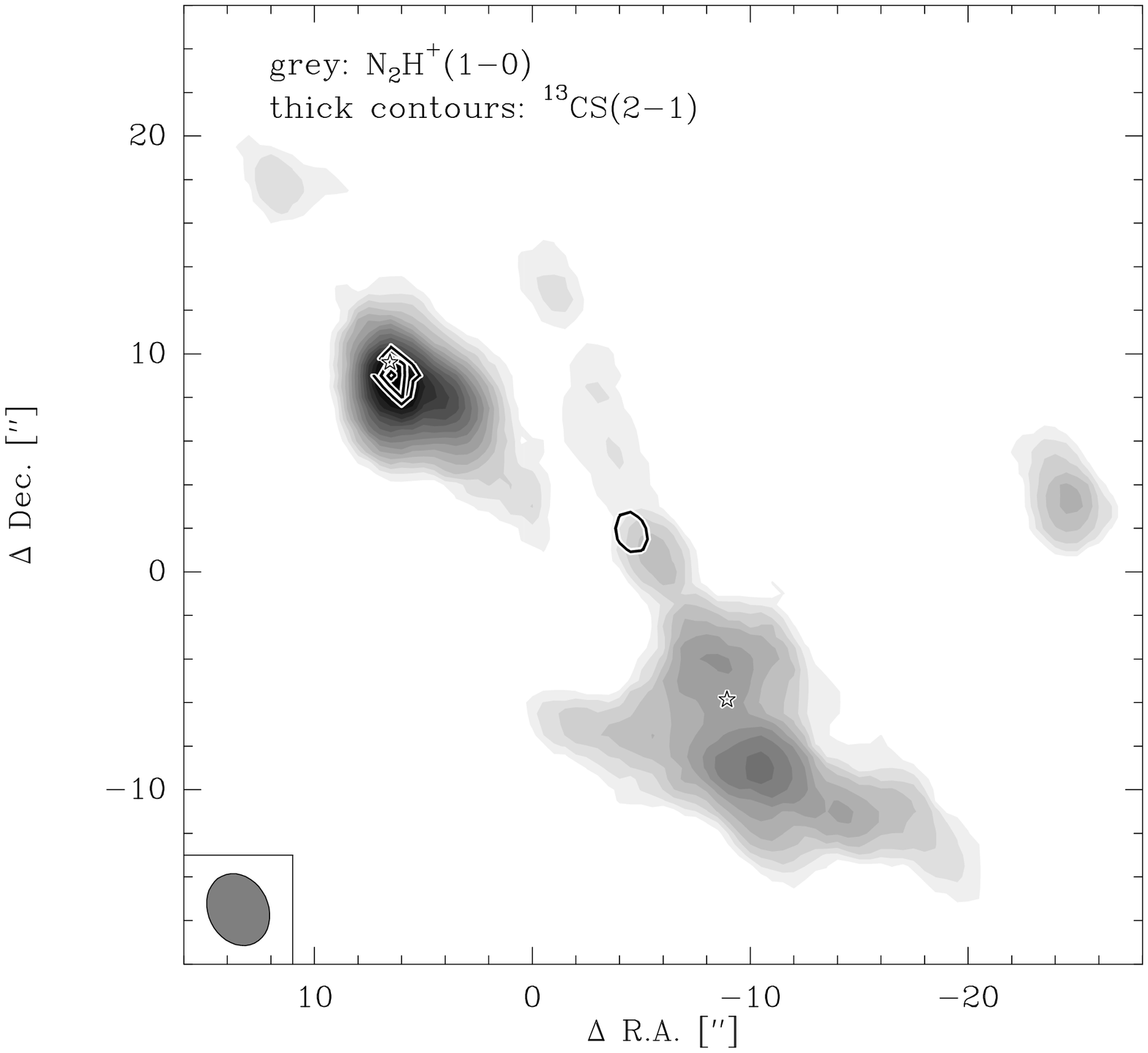}
\caption{The grey-scale shows the 0th moment or integrated intensity
  N$_2$H$^+$(1--0) map of the isolated N$_2$H$^+$(1--0) (F1,
  F=0,1$\rightarrow$1,2) hyperfine emission component as in Figure
  \ref{moments}. The thick contours present the $^{13}$CS(2--1)
  emission integrated from 6.9 to 9.5\,km\,s$^{-1}$ (0th moment). The
  contouring is done in $3\sigma$ steps of
  5\,mJy\,beam$^{-1}$\,km\,s$^{-1}$. The stars mark the positions of
  the main mm continuum peaks (Figs.~\ref{overview} \&
  \ref{continuum}), and the synthesized beam is shown at the
  bottom-left. The offsets are given with respect to the phase
  center.}
\label{13cs}
\end{figure}

In addition to the N$_2$H$^+$(1--0) emission, our spectral setup also
covered the $^{13}$CS(2--1) line. However, although detected, the
$^{13}$CS(2--1) emission is significantly weaker than
N$_2$H$^+$(1--0). Figure \ref{13cs} presents an overlay of the
integrated $^{13}$CS(2--1) with the N$_2$H$^+$(1--0) emission. It is
interesting to note that while the main mm and N$_2$H$^+$ peak is also
detected in the $^{13}$CS data, of the remaining N$_2$H$^+$ cores,
only peak 6 is associated with detectable $^{13}$CS emission above the
$3\sigma$ level. This already indicates peculiar low abundances
compared to more evolved regions, but likely also varying abundances
between the different cores.

Toward the main continuum and line peak, the integrated
$^{13}$CS(2--1) emission of 0.042\,Jy\,km\,s$^{-1}$ corresponds in the
Rayleigh-Jeans limit to 0.52\,K\,km\,s$^{-1}$.  How does that compare
to total single-dish fluxes? \citet{beuther2002a} observed the
CS(2--1) line toward that region with the IRAM\,30\,m telescope, and
the measured integrated intensity of 3.3\,K\,km\,s$^{-1}$ corresponds
to a flux of 17.6\,Jy\,km\,s$^{-1}$. Assuming a typical line intensity
ratio between $^{12}$CS and $^{13}$CS of $\sim$7 we get an integrated
$^{13}$CS intensity of 2.45\,Jy\,km\,s$^{-1}$ (toward the hot core
NGC6334I and the much younger sibling NGC6334I(N), the measured ratios
are 8 and 6.5, Walsh et al.~in prep.). Therefore, the missing flux we
loose by the interferometer observations amounts to approximately
80\%, similar to the $\sim$90\% for the mm continuum data (see section
\ref{cont}).

Assuming optically thin emission, we can use the $^{13}$CS(2--1) data
to estimate $^{13}$CS column densities of the compact gas component
following \citet{rohlfs2006} at the given low temperatures of
$\sim$15\,K.  Employing further a $^{12}$CS/$^{13}$CS ratio of 50, we
can estimate the main isotopologue's $^{12}$CS column densities and
derive the CS abundance by comparison with the H$_2$ column densities
calculated in section \ref{cont}. This results in a CS column density
of $\sim 1\times 10^{14}$\,cm$^{-2}$ and a CS abundance with respect
to H$_2$ of $\sim 5\times 10^{-11}$. Such very low values are
consistent with chemical models that predict CS column densities of
that order at the beginning of core evolution (e.g.,
\citealt{nomura2004}). The estimated abundance is about 2 orders of
magnitude lower than observed for example toward Orion-KL (e.g.,
\citealt{vandishoeck1998}), indicating CS depleting factors of $\sim
100$ in very young cores. For comparison, \citet{zhang2009} recently
found similarly high CO depletion factors (between 100 and 1000)
toward the IRDC G28.32.

\subsection{Virial masses}

Following \citet{maclaren1988}, we calculated the virial masses
$M_{\rm{vir}}=k_2\times R \times \Delta v^2$ with the constant $k_2$
between 190 and 126 for $1/r$ or $1/r^2$ density distributions, the
core radius $R$ (in units of parsecs) and the N$_2$H$^+$(1--0) line
width $\Delta v$. For the radius we assumed almost all sources to be
unresolved with an average N$_2$H$^+$(1--0) synthesized beam of $\sim
3.1''$, corresponding to the core diameter. Only peak 1 is resolved
and we used a radius of $3.5''$. Table \ref{n2h+} lists the range of
virial masses for the given line widths and assumed density
distributions.

The virial mass derived for the main core (between 7.5 and
11.4\,M$_{\odot}$) corresponds well to the gas mass derived above from
the mm continuum emission (see section \ref{cont}). This indicates
that the core may be in virial equilibrium.  Nevertheless, since this
core also exhibits the broadest line width, it is possible that the
core may be in a contracting phase prior to active star formation
(e.g., \citealt{caselli2002}). We doubt that the line broadening is
due to an early outflow since we did not detect any SiO emission in
previous observations \citep{beuther2007g}.  For most other N$_2$H$^+$
cores, we cannot draw conclusions regarding their virial status since
most of the virial masses are at or below our $3\sigma$ mass
sensitivity of $\sim$0.9\,M$_{\odot}$ derived from the continuum data.
Only the virial masses of the N$_2$H$^+$ cores 4, 10 and 12 are above
this $3\sigma$ sensitivity limit, indicating that at least these cores
are stable against gravitational collapse or may even be transient,
again expanding structures.

\section{Comparison with other star-forming regions}

How do the results of these two IRDCs compare with more typical
regions of low-mass starless cores, for example the Pipe nebula
\citep{alves2007,lada2008,rathborne2008b}? While some parameters like
the masses and the majority of cores being starless are similar
between the Pipe and the IRDCs discussed here, other parameters are
not. For example, the mean density of the Pipe cores is
$7\times10^3$\,cm$^{-3}$, approximately 2 orders of magnitude lower
than what we estimate for the IRDC\,19175 cores. This density
difference results also in different Jeans-length of the order 0.2\,pc
for the Pipe and $\sim$6000\,AU for the cores discussed here.
Interestingly, like in our case where the observed core separation
corresponds well to the Jeans-length, the same is the case for the
Pipe nebula with a much wider average core separation of
$\sim$0.26\,pc (Joao Alves, private communication). This supports the
long-standing argument that the relatively simple Jeans-analysis can
be considered a good approximation for the expected fragmentation
scales. But why are the average densities so different? For a more
conclusive answer to this question, a larger statistical sample would
be required, however, the environments of the two regions already
indicate a possible reason. While the Pipe nebula is relatively
isolated, the IRDCs discussed in this paper are close to the
intermediate- to high-mass star-forming region IRAS\,19175+1357.
Therefore, it is likely that the environment -- either direct
influence from the more evolved IRAS source and/or the density and
column density structure of the original gas clump -- influences the
density structure of the nearby cores and this way also determines the
Jeans- and fragmentation length of the neighboring next generation of
star formation.

The observed line width in typical low-mass starless cores is usually
found to be close to thermal where the non-thermal line width
contribution rarely exceeds the thermal part (e.g.,
\citealt{andre2007,rathborne2008b,foster2009}). The situation is
slightly different in starless cores associated with high-mass
star-forming regions where on average broader line widths are found.
For example, \citet{fontani2008} found N$_2$H$^+$ line width around
0.7\,km\,s$^{-1}$ in the intermediate- to high-mass star-forming
region IRAS\,05345+3157, whereas \citet{wang2008} and
\citet{zhang2009} found NH$_3$ line widths around 1.7\,km\,s$^{-1}$
towards the youngest massive core associated with G28.34+0.06.

The average line width found in the 17 cores of IRDC\,19175-4 and
IRDC-19175-5 is $\sim$0.8\,km\,s$^{-1}$ with a spread between 0.5 and
1.8\,km\,s$^{-1}$. As discussed in section \ref{n2h+_main}, the lower
end of that regime is less accurately determined since our velocity
resolution is only 0.2\,km\,s$^{-1}$, barely resolving the narrowest
lines. Similar to \citet{fontani2008}, our measured line widths exceed
those of the low-mass cores, however, they are still narrower than
observed toward more evolved massive star-forming regions (e.g.,
high-mass protostellar objects, HMPOs, or ultracompact H{\sc ii}
regions, UCH{\sc ii}s,
\citealt{beuther2002a,churchwell1990,cesaroni1991}). This indicates a
lower level of turbulent motions at the onset of intermediate-mass
star formation compared to its successive evolution. Furthermore, the
many N$_2$H$^{^+}$ sub-sources show that such intermediate-mass cores
can fragment significantly and potentially form small groups or
clusters.

Maybe a bit surprising, the line widths found toward the very young
clump in G28.34+0.06 by \citet{wang2008} are intermediate between the
narrower lines we and \citet{fontani2008} observed, whereas they are
still less broad than than those of typical HMPOs or UCH{\sc ii}
regions. It is interesting to compare a few other quantities of this
particular region P1 in G28.34+0.06 (see \citealt{zhang2009}): The
average separation between their 5 sub-cores is 0.19\,pc, similar to
the Pipe and different to the case discussed here. In contrast to
that, the average densities in G28.34+0.06 range between $10^6$ and
$10^7$\,cm$^{-3}$, comparable to the values found in IRDC\,19175-4 and
IRDC-19175-5. As outlined above, for the Pipe nebula, densities,
Jeans-length and observed core separation match rather well. This is
not the case for the P1 region in G28.34+0.06.  Their densities imply
smaller Jeans length and lower Jeans-masses than observed. As
discussed by \citet{zhang2009}, in the case of G28.34+0.06, not only
thermal pressure and gravity seem to control the fragmentation of the
initial gas clump, but other processes are required. The broader
observed line width there indicates a significant turbulent
contribution which may help to stabilize the cloud cores against
further fragmentation at that evolutionary stage.
  
The situation for IRDC\,19175-4 and IRDC-19175-5 is different to
G28.34+0.06. Since here densities, Jeans-lengths and observed core
separation match well, turbulence seems to play a less important role
for the fragmentation processes. The cloud sub-structure can well be
dominated by the interplay of mainly thermal pressure and gravity.

\section{Discussion and Conclusions}

Combining Spitzer IRAC and MIPS data from 3.6 to 70\,$\mu$m with mm
continuum and N$_2$H$^+$ observations from the Plateau de Bure
Interferometer and the IRAM 30\,m telescope, we identify two very
young gas and dust cores of intermediate mass ($\sim$87\,M$_{\odot}$
together for IRDC\,19175-4 and IRDC\,19175-5) in the vicinity of the
intermediate- to high-mass star-forming region IRAS\,19175+1357. The
high-spatial-resolution PdBI 3.23\,mm continuum data clearly resolve
IRDC\,19175-4. However, IRDC\,19175-5 is only barely detected above
the $3\sigma$ level. More importantly, the N$_2$H$^+$(1--0) line
reveals a completely different structure with 17 separate emission
sources within the vicinity of IRDC\,19175-4 and IRDC\,19175-5.
Fitting the full N$_2$H$^+$(1--0) hyperfine structure, we are able to
derive the line parameters (intensity, peak velocity, line width and
optical depth) for all cores, and from that the N$_2$H$^+$ column
densities and virial masses.

Of particular interest are the measured line width of the N$_2$H$^+$
cores. Only the most massive core IRDC\,19175-4 with a mass of
$\sim$10\,M$_{\odot}$ exhibits a line width of $\sim$1.8\,km\,s$^{-1}$
whereas all other N$_2$H$^+$ cores have significantly narrower line
widths below or around 1\,km\,s$^{-1}$.  While the measured line
widths between 0.5 and 1.8\,km\,s$^{-1}$ are still larger than the
thermal value of $\sim$0.15\,km\,s$^{-1}$ (at 15\,K), the line widths
are narrow compared to more evolved high-mass star-forming regions
(e.g., \citealt{churchwell1990,hatchell1998b,beuther2002a}) and hence
indicate a low level of internal turbulence. Matching densities,
Jeans-lengths and observed core separation indicate that the
fragmentation process may be dominated by the interplay of mainly
thermal pressure and gravity, but not so much by turbulence.  These
low-turbulence cores are in the vicinity of a high-mass star-forming
region with approximately $9\times 10^4$\,AU separation from
IRAS\,19175+1357. As outlined in section \ref{intro}, although we
cannot proof direct interaction between the more evolved IRAS source
and the younger IRDCs, both should stem from the same gas clump that
originally fragmented and produced (massive) star-forming regions of
different evolutionary stages.

A comparison of the N$_2$H$^+$ column densities with the corresponding
H$_2$ values (and upper limits) derived from the dust continuum
emission, indicates significant abundance differences between the
N$_2$H$^+$ cores. While most of the lower-mass cores are consistent
with typical N$_2$H$^+$ abundances of the order $10^{-10}$ with
respect to H$_2$, the main core IRDC\,19175-4 has lower abundances of
the order $10^{-11}$. Since this source also shows the broadest line
width, we interprete this lowered abundances as likely due to higher
densities from the contracting core which increases the recombination
rate and hence lowers the abundances of the ions. Furthermore, the
weak $^{13}$CS(2-1) detection allows us to estimate CS column
densities and abundances, and we find CS depletion factors of the
order 100 compared to more evolved regions like Orion-KL.

The virial analysis and line width measurements indicate that the main
core IRDC\,19175-4, although being close to virial equilibrium, may
already be in a phase of core contraction and hence at the verge of
star formation. For the remaining cores, the status can be different.
Since we have only upper limits for the gas masses from the dust
continuum emission, a real virial analysis is difficult, however,
these upper limits are consistent with virial stable cores. Similarly,
the separation between the different sub-cores corresponds
approximately to the Jeans-length at 15\,K and a density of
$10^6$\,cm$^{-3}$. Hence, the cores can be stable against further
fragmentation. All this indicates that many of them may never enter
star formation activity but could rather be transient or stable cores
in the vicinity of other massive star-forming regions.

In summary, here we report the characterization of 17 low-to
intermediate mass N$_2$H$^+$ cores with very low levels of turbulence
in the vicinity of a high-mass star-forming region. The non-detection
up to 70\,$\mu$m wavelength, the low levels of turbulence as well as
no other signs of star formation in previous studies indicates that we
are dealing with starless cores. While some of them maybe contracting
and may start active star formation in the future, many of them may be
stable against collapse and hence potentially even only transient
structures. The average densities are approximately 2 orders of
magnitude higher than in the Pipe nebula which is also manifested in
smaller core separations. We note that these rather quiescent cores
are spatially associated with more turbulent intermediate- to
high-mass star-forming cores, implying that low and high turbulence
regions can co-exist in relatively close proximity.

\begin{acknowledgements} 
  Thanks a lot to Helmut Dannerbauer and Roy van Boekel for helping
  with the Spitzer data. Furthermore, we like to thank C.~de Vries for
  providing the N$_2$H$^+$ column density calculation routine. In
  addition, we appreciated a lot the referee's comments improving the
  paper.  H.B.~acknowledges financial support by the
  Emmy-Noether-Program of the Deutsche Forschungsgemeinschaft (DFG,
  grant BE2578).
\end{acknowledgements}


\end{document}